\documentclass[%
 reprint,
superscriptaddress,
 amsmath,amssymb,
 aps,
 pra,
]{revtex4-1}

\usepackage{graphicx}% Include figure files
\usepackage{dcolumn}% Align table columns on decimal point
\usepackage{bm}% bold math
\begin{document}

\preprint{APS/123-QED}

\title{A Gauge Invariant Formulation of Interband and Intraband Currents in Solids}

\author{Guilmot Ernotte}
\affiliation{Joint Attosecond Science Laboratory, National Research Council of Canada and University of Ottawa, Ottawa, Ontario K1A 0R6, Canada}

\author{T. J. Hammond}
%\affiliation{Joint Attosecond Science Laboratory, National Research Council of Canada and University of Ottawa, Ottawa, Ontario K1A 0R6, Canada}
\affiliation{Department of Physics, University of Windsor,
Windsor, Ontario N9B 3P4, Canada}

\author{Marco Taucer}
\email{Marco.Taucer@nrc-cnrc.gc.ca}
\affiliation{Joint Attosecond Science Laboratory, National Research Council of Canada and University of Ottawa, Ottawa, Ontario K1A 0R6, Canada}

\date{\today}

\begin{abstract}
Experiments and simulations in solid-state high harmonic generation often make use of the distinction between interband and intraband currents. These two contributions to the total current have been associated with qualitatively different processes, as well as physically measurable signatures, for example in the spectral phase of harmonic emission. However, it was recently argued [P. F\"{o}ldi, Phys. Rev. B \textbf{96}, 035112 (2017)] that these quantities can depend on the gauge employed in calculations. Since physical quantities are expected to have gauge-independent values, this raises the question of whether the decomposition of the total current into interband and intraband contributions is physically meaningful, or merely a feature of a particular mathematical representation of nature. In this article, we explore this apparent ambiguity. We show that a closely related issue arises when calculating instantaneous band populations. In both the case of inter/intraband currents and in the case of instantaneous band populations, we 
propose definitions which are gauge-invariant, and thus
allow these quantities to be calculated consistently in any gauge.
\end{abstract}
\maketitle

\section{Introduction}

The strong field of a pulsed laser can drive extremely nonlinear currents in a solid, leading to the emission of high-order harmonics of the fundamental frequency that can span the visible spectrum and extend into the extreme ultraviolet \cite{Ghimire2011}. The process can take place in a wide range of materials, from dielectrics to semiconductors to semimetals, and can leave the material undamaged \cite{Schubert2014,Liu2017a,Sivis2017,Yoshikawa2017,Luu2015,Vampa2015}. Strong field and attosecond science in condensed matter is an extension of the long-standing field of high harmonic generation, which was for many years confined to gas-phase atoms and molecules \cite{Krausz2009}. Solids remain a new area of this field, in which some basic questions remain unanswered, while others may not even be precisely defined.

In that respect, as part of the search for an underlying physical picture of the harmonic generation process, experimentalists and theorists alike have focused much attention on the division of the total current into interband and intraband processes. This conceptual separation is appealing in part because the interband picture bears a strong similarity to the well-understood gas-phase model \cite{Corkum1993,Vampa2015b,Vampa2015}, while the intraband picture is qualitatively different and for the most part unique to the solid state \cite{Luu2015,Garg2016}. Experiments have access to the complex amplitude of the emitted harmonics which reflects the coherent sum of the interband and intraband contributions, whatever their relative weight may be. For now, a clean separation is only possible in calculations; any conclusions about the dominance of one mechanism or the other relies on a comparison with theoretical predictions of the interband and intraband spectra \cite{Vampa2015b,Vampa2015,Vampa2014,Luu2015,Garg2016, Hammond2017,Schubert2014,Hohenleutner2015,Taucer2017}. Recently, however, F\"{o}ldi showed that this separation may be gauge-dependent \cite{Foldi2017}. That is, a different choice of the gauge, which should leave all physical quantities unchanged, leads to different values for the interband and intraband currents. However, the total current, which relates to the experimentally observed harmonic spectrum, is not gauge-dependent. This raises the question of whether this conceptual decomposition of the current is physically meaningful. Is the interband current an observable?

Other quantities may also be easy to calculate, but hard to access in experiments. An example is the instantaneous band population. In the strong-field physics of solids, simulations often show a transient conduction band population which oscillates with the applied field, but mostly returns to the valence band at the end of the pulse. The fraction of the population which remains in the conduction band at the end of the pulse depends on the band structure and the pulse shape, as well as the dephasing time constant \cite{McDonald2017}. 
%Below, we will show that the instantaneous population can be gauge-dependent in precisely the same way as the separation of currents into their inter- and intra-band components. 
While this final population is gauge-independent, the transient population dynamics during the laser pulse's illumination can be subject to a gauge-dependence that is analogous to that of the inter- and intra-band currents, as we will show below.
This raises the question of the significance of the instantaneous band populations. Can such a quantity be precisely defined, particularly given that in the presence of a strong driving field the instantaneous eigenstates are the dressed states, which are not the same as the field-free eigenstates? The question is all the more compelling given that some attosecond-probe experiments appear to measure precisely this quantity \cite{Sommer2016}.

The aim of the present article is to create gauge-invariant definitions of these quantities of interest. We start by identifying the Hermitian operators corresponding to the instantaneous band populations, and to the interband and intraband currents. Once defined, we derive their gauge-dependent transformations, which ensures gauge-invariant physical predictions. While we primarily focus on the commonly used velocity and length gauges, our definition is equally valid in any other gauge.

This paper is organized as follows. In Section \ref{sec:Approach}, we introduce the theoretical formalism and the details of our numerical calculations. Section \ref{sec:NaiveDefinitions} discusses an intuitive, but problematic, approach to defining interband and intraband currents as well as band populations. 
In Section \ref{sec:populations}, we provide a more rigorous definition of the instantaneous band population, as a Hermitian operator. We then calculate the gauge-transformation of its matrix elements. Section \ref{sec:currents} similarly describes the interband and intraband currents in terms of Hermitian operators, with corresponding gauge transformations. The improved definitions of these quantities yield gauge-invariant predictions. Finally, in Section \ref{sec:Discussion}, we show that our definitions give reasonable physical descriptions, and we discuss the choices made in coming to this formulation.

\section{Theoretical Approach}
\label{sec:Approach}
For a single particle in a one-dimensional periodic potential, $V_0(x+a_0)=V_0(x)$ with a lattice constant of $a_0$, the Hamiltonian in the absence of the laser field is
\begin{equation}
\hat{H}_0 = \frac{\hat{p}^2}{2} + V_0(\hat{x}).
\label{eqn:FieldFreeHamiltonian}
\end{equation}
Here and throughout this paper we use atomic units, except where other units are specified. The eigenstates of this Hamiltonian can be labeled by a band index, $n$, and the crystal momentum, $k$:
\begin{equation}
\hat{H}_0 | \phi_{nk} \rangle = \varepsilon_{n}(k) | \phi_{nk} \rangle.
\end{equation}
The energies, $\varepsilon_{n}(k)$, trace out the band structure, and the Bloch functions, expressed in the position basis, have the property $\langle x|\phi_{nk} \rangle \equiv \phi_{nk}(x) = \sqrt{\frac{a_0}{2 \pi}} e^{ikx} u_{nk}(x)$, where $u_{nk}(x+a_0)=u_{nk}(x)$ is periodic and normalized over one unit cell.

\begin{figure}
\includegraphics{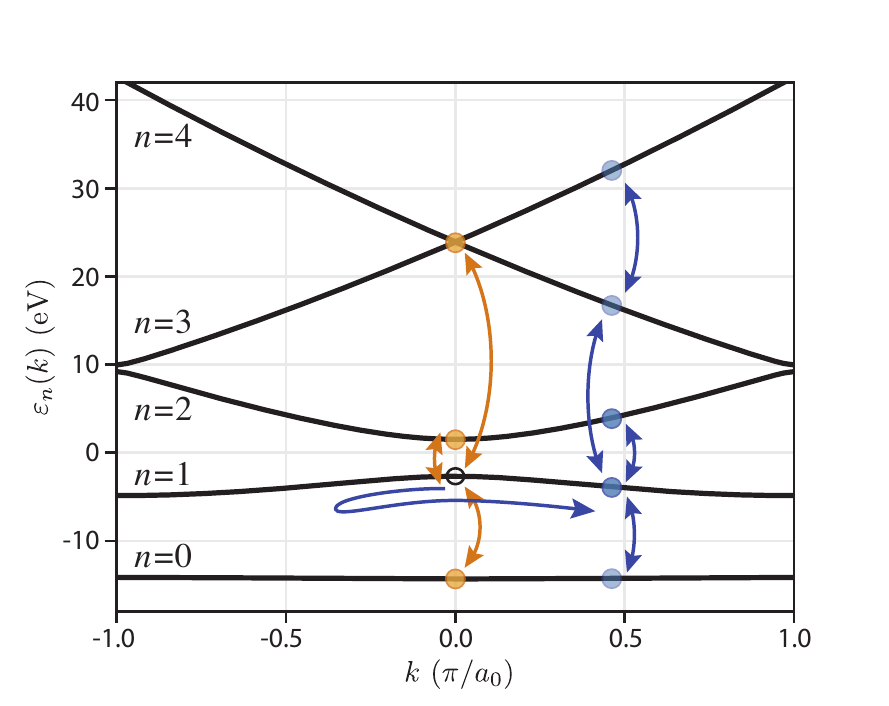}
\caption{Band structure for the Mathieu potential,
showing the lowest five bands, labeled with their band
indices, as a function of crystal momentum in the first
Brillouin zone. The black circle in band 1 shows the initial
condition for our simulation. Orange and blue arrows/circles illustrate the dynamics as described in the velocity and length gauges, respectively. Both gauges describe the time-dependent wavefunction as a superposition of Bloch states at a particular crystal momentum, but in the velocity gauge the crystal momentum is fixed, while in the length gauge it oscillates with the vector potential.
% (b) Magnitude of the interband
% momentum matrix elements between band 1 and all
% the other bands. (c) Squared modulus of the functions, $\Delta_{mn}(0,k)$ for $m=1$.
% Roughly speaking, these indicate the degree to which
% each periodic function, at k=0, is required to represent a
% wavefunction with wavevector k in band n.
}
\label{fig:1}
\end{figure}

As a model system, we use the previously studied Mathieu potential, $V_0(x) = -V_0 \left[ 1 + \cos(2 \pi x / a_0) \right]$, with $V_0 = 0.37$ and $a_0 = 8$ atomic units \cite{Wu2015,Wu2016,Liu2017,Ikemachi2017,Liu2017b}. We solve the Time-Independent Schrodinger Equation (TISE) in the position basis, with periodic boundary conditions, to find the field-free eigenstates (Bloch states). These are then used as the basis for calculations of the time dynamics in a driving laser field. Figure \ref{fig:1} shows the band structure as a function of crystal momentum for the first five bands. The black circle in the center of the band with index $n=1$ represents the initial condition for simulations: $| \phi_{1,k=0} \rangle$, a single electron at the $\Gamma$-point in band-1. Roughly speaking, bands 1 and 2 can be thought of as the valence band and the first conduction band, respectively. However, our calculation considers band-0 to be unoccupied, as well as all other $k$-points in band-1. While this simplification does not represent the reality of valence bands, it allows a comparison with previous reports and has no effect on the conclusions of this work. 

The effect of a laser field, which we treat here within the dipole approximation, leaves freedom with respect to the gauge chosen, since the field can be divided non-uniquely between a scalar potential, $\Phi$, and a vector potential, $A$. In the length gauge, the field is incorporated exclusively through the scalar potential,
\begin{eqnarray}
\Phi^{(l)}(x,t)&=& - F(t)\cdot x  \nonumber \\
A^{(l)}(t)&=&0 ,
\end{eqnarray}
where $F(t)$ represents the electric field. In the velocity gauge, the situation is reversed,
\begin{eqnarray}
\Phi^{(v)}(x,t)&=& 0  \nonumber \\
A^{(v)}(t) &=&  -\int_{-\infty}^t F(t') dt'.
\end{eqnarray}
The Hamiltonian, including the interaction with the light field, under the dipole approximation, can be written in any gauge, labelled with superscript $(g)$, as
\begin{equation}
\hat{H}^{(g)}=\frac{1}{2} \left[\hat{p}+A^{(g)}(t) \right]^2 + V_0(\hat{x}) - \Phi^{(g)}(\hat{x},t) .
\label{eqn:GaugeHamiltonian}
\end{equation}
Note that when the field and the (velocity gauge) vector potential are both zero (that is, before or after the pulse), the Hamiltonian reduces to $\hat{H}_0$ in both the velocity and length gauge. 

Hermitian operators corresponding to observables transform between the two gauges according to \cite{Han2010,Bandrauk2013,NOTEoperators},
\begin{equation}
\hat{O}^{(v)}=e^{-iA^{(v)} \hat{x}} \hat{O}^{(l)} e^{iA^{(v)} \hat{x}},
\label{eqn:OperatorTransformation}
\end{equation}
and the wavefunctions are related by
\begin{equation}
| \psi^{(v)} \rangle = e^{-iA^{(v)} \hat{x}} |\psi^{(l)} \rangle .
\label{eqn:WavefunctionTransformation}
\end{equation}
%More generally, wavefunctions and operators, in gauge $(g)$, are transformed by the unitary operator, 
More generally, wavefunctions and operators are transformed from gauge $(g_1)$ to gauge $(g_2)$ by the unitary operator %$e^{i[A^{(g_2)}\hat{x} - A^{(g_1)}\hat{x}]}$. 
\begin{equation}
\hat{U}_{(g_1)\rightarrow (g_2)} \equiv e^{i[A^{(g_2)} - A^{(g_1)}]\hat{x}} .
\label{eqn:GeneralGaugeTransformation}
\end{equation}
%The approach we take in this article starts from the length gauge, whose vector potential is zero. 
In this article, we choose to equate the field-free operator for our quantities of interest with the operator's representation in the length gauge. While this is not the only possible choice, we will provide a justification below, and show that this definition gives reasonable physical predictions.  
Starting from the length gauge, then, the transformation to any other gauge, $(g)$, is described by 
\begin{equation}
\hat{U} \equiv \hat{U}_{(l)\rightarrow (g)} = e^{iA^{(g)} \hat{x}} .
\label{eqn:UnitaryOperator}
\end{equation}
In each gauge, the vector potential dictates a time-dependent transformation. As long as wavefunctions and Hermitian operators transform via the unitary operator of Eq. \ref{eqn:UnitaryOperator}, expectation values remain unchanged by the gauge transformation. In other words, the different gauges all represent the same physics.

We solve the Time-Dependent Schr\"{o}dinger Equation (TDSE) in the field-free basis numerically. We write wavefunctions in gauge $g$ as
\begin{equation}
|\psi^{(g)} (t) \rangle = \sum_{n} \int_{BZ} dk ~ c_{nk}^{(g)}(t) | \phi_{nk} \rangle.
\label{eqn:WavefunctionExpansion}
\end{equation}
The amplitudes, $c_{nk}^{(g)}(t)$, depend on the gauge. For numerical simulations, the laser field is defined by $A^{(v)}(t) = A_0 \cos^4(\omega_0 t / 2 n_c) \cos(\omega_0 t)$, where $A_0 = 0.3$ is the peak vector potential, $n_c = 11$ is the number of cycles in the pulse, and a fundamental frequency of $\omega_0 = 2 \pi c / \lambda$, with $c$ the speed of light and $\lambda = 3.2~\mu\rm{m}$ the wavelength. Our parameters are identical to those used by Wu \textit{et al.} \cite{Wu2015,Wu2016}. 

%***
%The time-dependence of the coefficients, $c_{nk}^{(g)}(t)$, can be expressed as a set of coupled differential equations, which take a different form in each gauge. In the velocity gauge, 
% \begin{equation}
% i \frac{\partial}{\partial t} c_{nk}^{(v)} = 
% \left[ \varepsilon_{n}(k) + \frac{1}{2} {A^{(v)}}^2 \right] c_{nk}^{(v)} + A^{(v)} \sum_{n'} p_{nn'}(k) c_{n'k} . 
% \end{equation}
% We can define the matrix elements of momentup as $\langle \phi_{nk} | \hat{p} | \phi_{mk'} \rangle \equiv p_{nm}(k) \delta(k-k') $
% ***

In the presence of the laser field, the Hamiltonian expressed in the field-free basis acquires off-diagonal elements that couple the Bloch states. This coupling is determined by the matrix elements of the momentum operator in the velocity gauge, or by the matrix elements of the position operator in the length gauge. 
The resulting set of coupled differential equations for the coefficients, $c_{nk}^{(g)}(t)$, takes a different form in each gauge. In the velocity gauge, 
\begin{equation}
i \frac{\partial}{\partial t} c_{nk}^{(v)} = 
\left[ \varepsilon_{n}(k) + \frac{1}{2} {A^{(v)}}^2 \right] c_{nk}^{(v)} + A^{(v)} \sum_{n'} p_{nn'}(k) c_{n'k}^{(v)} . 
\end{equation}
The momentum operator only couples states with the same $k$-value, $\langle \phi_{nk} | \hat{p} | \phi_{n'k'} \rangle \equiv p_{nn'}(k) \delta(k-k') $, meaning that the initial crystal momentum, in the velocity gauge description, remains constant even under the influence of the laser field. In the length gauge, 
\begin{equation}
i \frac{\partial}{\partial t} c_{nk}^{(l)} = 
\varepsilon_{n}(k) c_{nk}^{(l)} + i F \frac{\partial}{\partial k} c_{nk}^{(l)} + F \sum_{n'} \xi_{nn'}(k) c_{n'k}^{(l)} . 
\end{equation}
The position matrix elements likewise contain a part which mixes states of the same $k$-value, denoted $\xi_{nn'}(k)$, which, for 
%$n\neq n'$, 
non-degenerate states,
is given by $\xi_{nn'}(k) = -i p_{nn'}(k)/(\varepsilon_n(k) - \varepsilon_{n'}(k) )$. However, the position operator additionally contains a differential term which couples neighbouring $k$-values, and leads to the acceleration theorem: a state initially having $k=k_0$ evolves into states with $k(t)=k_0+A^{(v)}(t)$. The acceleration theorem illustrates an important difference between these two gauges: in the velocity gauge, the crystal momentum remains fixed, while in the length gauge, the evolution of the wavefunction in the laser field leads to an oscillation of the electron's crystal momentum. This is illustrated by the coloured circles and arrows in Fig. \ref{fig:1}. %It is possible to connect these two pictures using overlap functions that will be defined later.
%The connection between these two representations involves the functions, $\Delta_{mn}(k_1,k_2)$, shown in Fig. \ref{fig:1}c, which will be defined later.

\section{Definitions Based on Band Indices}
\label{sec:NaiveDefinitions}
In this section we discuss an intuitive yet problematic procedure for defining band populations and inter/intraband currents, in which the band indices of coefficients are used to identify  band-dependent quantities. This procedure is known to give gauge-dependent results \cite{Foldi2017}. In subsequent sections, we will attempt to improve upon these definitions.

\begin{figure}
\includegraphics{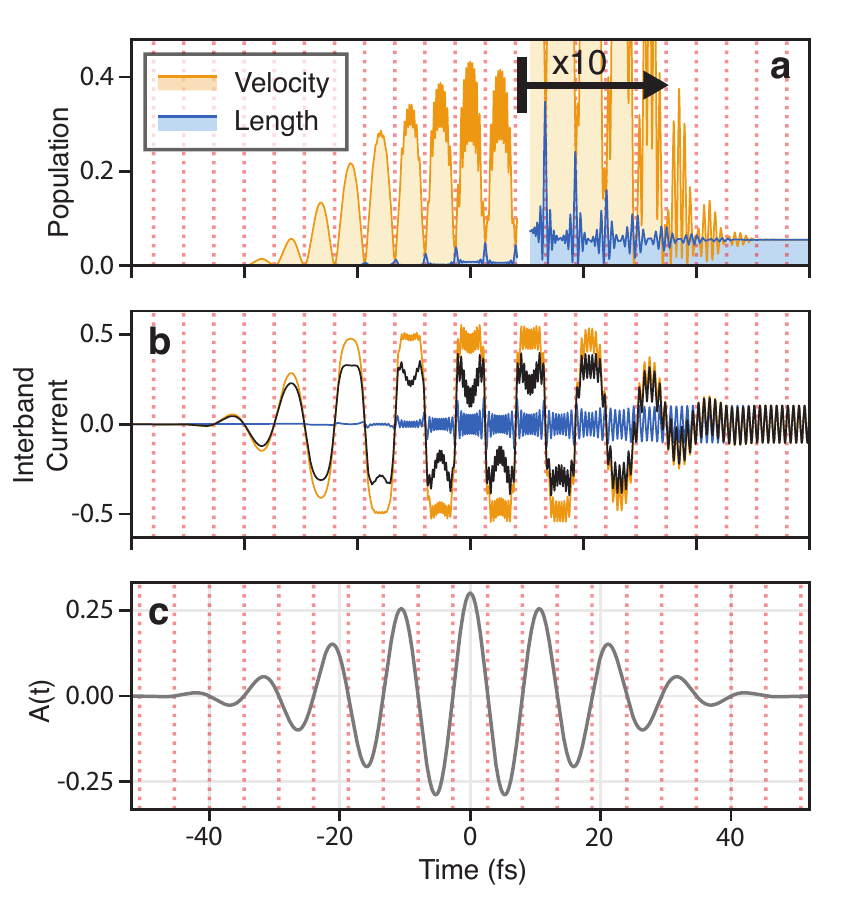}
\caption{Dynamics calculated using definitions based on band indices. (a) Conduction band population, defined as the
squared modulus of the time-dependent coefficients, calculated in the velocity and length gauges. The region on the right side is magnified in the
vertical dimension by a factor of ten. (b) Interband current,
defined as the second summation in Eqn. \ref{eqn:NaiveCurrent}, calculated in the velocity gauge
and length gauge. The total current, which is gauge-independent
provided one makes the transformation $\hat{p}^{(g)} =
\hat{p}+A^{(g)}$, is shown in black for reference. (c) Velocity gauge vector potential as a
function of time. Vertical dashed lines in all plots indicate
the zeros of this vector potential.}
\label{fig:2}
\end{figure}

Solving the Schr\"{o}dinger equation in either gauge gives the time-dependent coefficients of the basis states, and thereby the wavefunction. In particular, the coefficients, $c_{nk}^{(g)}(t)$, are associated with the state $| \phi_{nk} \rangle$, and, according to the Born rule, their modulus squared seems to represent the probability of finding an electron in that particular state. 
This reasoning implies that the instantaneous band population in band $n$ is $\int dk |c_{nk}^{(g)}|^2$. The resulting instantaneous conduction band population is shown in Fig. \ref{fig:2}a, for the velocity gauge (orange) and length gauge (blue). 
Both gauges describe a transient population which oscillates and nearly completely returns to the valence band at the end of the pulse. 
Both calculations agree on the final population that remains in the conduction band at the end of the pulse.  However, in the velocity gauge, the transient conduction band population is much larger, as much as 40\%, and it is peaked at the (velocity gauge) vector potential maxima, whereas in the length gauge the apparent conduction band population is peaked at the vector potential zeros. Also, whenever the vector potential (Fig. \ref{fig:2}c) is zero, indicated by vertical dashed lines, the two gauges agree. This is to be expected since the unitary transformation (Eq. \ref{eqn:UnitaryOperator}) is the identity in that case. But in general, the two calculations provide very different pictures of the conduction band population, and whenever the vector potential is non-zero, it is not clear which one to trust. In the recent literature, it appears that both quantities have been reported \cite{Sommer2016,Virk2007,Tancogne-DeJean2017,Schultze2013,Schlaepfer2018}. It would be desirable to find a gauge-independent formulation of the instantaneous band population.

Turning now to the current, we can express it simply in terms of the kinematical momentum, and expand its expectation value in terms of Bloch states,
\begin{eqnarray}
j(t) &=&  - \langle \psi^{(g)}(t) | \hat{p}_{kin}^{(g)} | \psi^{(g)}(t) \rangle \nonumber \\
&=& 
-\sum_{n} \int_{BZ}dk |c_{nk}^{(g)}|^2 \langle \phi_{nk} | \hat{p}_{kin}^{(g)} | \phi_{nk} \rangle \nonumber \\
&& - \sum_{n, n'\neq n} \int_{BZ}dk c_{nk}^{*(g)} c_{n'k}^{(g)} \langle \phi_{nk} | \hat{p}_{kin}^{(g)} | \phi_{n'k} \rangle ,
\label{eqn:NaiveCurrent}
\end{eqnarray}
noting that in the velocity gauge the kinematical momentum is $\hat{p}_{kin}^{(g)} = \hat{p} + A^{(g)}$, where $\hat{p}$ refers to the \textit{canonical} momentum. The summation over all basis states has been split into a summation over terms involving the same band indices (intraband) and a summation over terms involving different band indices (interband). Figure \ref{fig:2}b shows the interband current, so defined, as calculated in the velocity (orange) and length (blue) gauges. As in the case of the instantaneous conduction band population, the two calculations give very different results. Importantly, the total current, shown in black, is gauge-independent as long as the kinematical momentum is used. Since the total current is what gives rise to the measured harmonic spectrum, there is no question that it is a physically meaningful quantity, and its gauge-invariance is expected. However, as long as the treatment of interband and intraband currents depends on the gauge, it is not clear that these are physical quantities.

%F\"{o}ldi's recent work showed that the total current is a gauge-independent quantity, as expected [cite Han and Madsen], but the decomposition into the summations shown in Eq. \ref{eqn:NaiveCurrent} is gauge-dependent. 

\section{Band Populations}
\label{sec:populations}
In this section we propose a gauge-invariant definition of the instantaneous band populations. Our approach is to identify its corresponding Hermitian operator. 
%We then have at our disposal the transformation rules outlined above.
We then transform it with the proper unitary operators to make it gauge-independent.
Before proceeding, we consider the general question of the population of \textit{any} quantum state, $| S \rangle$. 
The probability to find a system in this state is evidently the squared modulus of the projection of the system's state onto $| S \rangle$. Another way to say this is that it is the expectation value of a Hermitian operator (an observable), namely the projection operator, $\hat{\Pi}_S \equiv | S \rangle \langle S |$.

Likewise the operator representing band population is an operator that projects onto all the states within a given band, $m$,
\begin{equation}
\hat{\Pi}_m = \int_{BZ}dq \hat{\Pi}_{mq} \equiv \int_{BZ}dq | \phi_{mq} \rangle \langle \phi_{mq} | ,
\label{eqn:BandProjection}
\end{equation}
where $\Pi_{mq}$ is the projection onto a single eigenstate, $| \phi_{mq} \rangle$, of the field-free Hamiltonian. Since this operator was defined without reference to a gauge, nor a field, we refer to it as a ``field-free operator''.
The expectation value of $\hat{\Pi}_m$ represents the band population, and Eq. \ref{eqn:UnitaryOperator} provides its transformation. 
%In the length gauge, this transformation is trivial, and we find that the band population is given by $|c_{mq}^{(l)}(t)|^2$, as shown in Fig. \ref{fig:2}a. However, the situation is not as simple in the velocity gauge. The operator representing this same quantity must be written
In a gauge $(g)$ the operator transforms as
\begin{equation}
\hat{\Pi}_{mq}^{(g)} = e^{-iA^{(g)} \hat{x}} \hat{\Pi}_{mq} e^{iA^{(g)} \hat{x}},
\label{eqn:InvariantBlochProjection}
\end{equation}
and its matrix elements, needed to compute the expectation value, in the field-free basis are
\begin{eqnarray}
 &\langle \phi_{nk} |&  \hat{\Pi}_{mq}^{(g)} | \phi_{n'k'} \rangle \nonumber \\
&&=
\langle \phi_{nk} | e^{-iA^{(g)} \hat{x}} |\phi_{mq} \rangle \langle \phi_{mq} | e^{iA^{(g)} \hat{x}} |\phi_{n'k'} \rangle
\nonumber \\
&&= U_{mn}^{\dagger}(q,k) U_{mn'}(q,k')
.
\end{eqnarray}

\begin{figure}%[b!]
\includegraphics{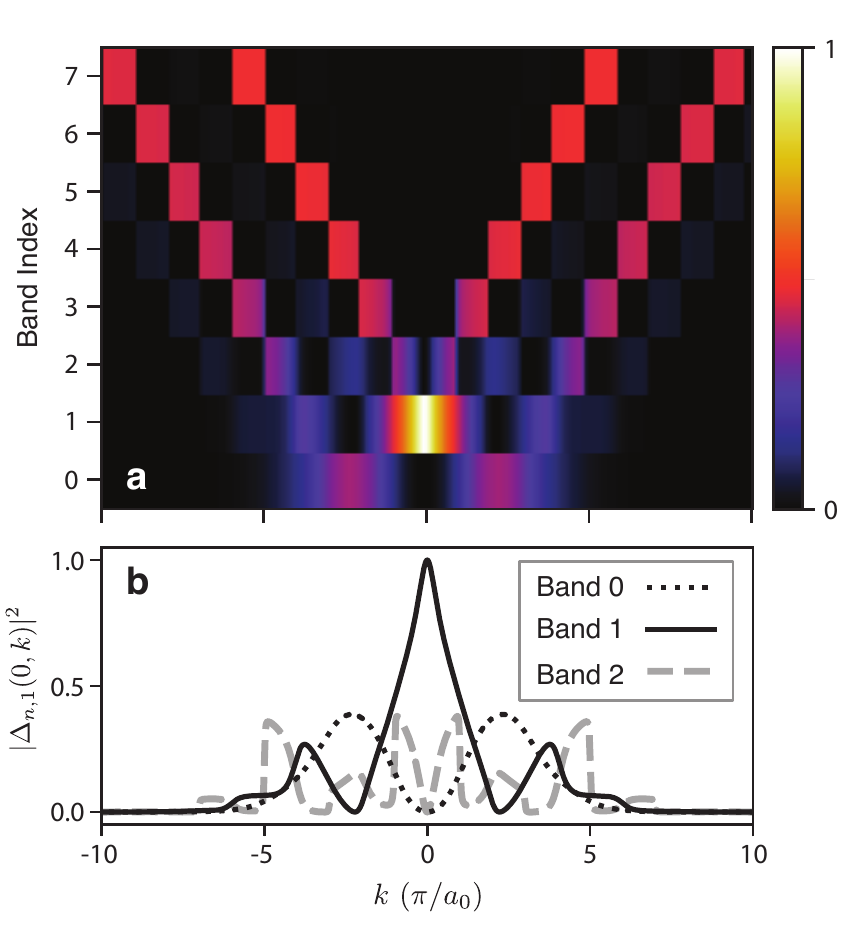}
\caption{Squared modulus of the functions, $\Delta_{n,1}(0,k)$, plotted over ten Brillouin zones of reciprocal space, for (a) the lowest seven bands and (b) for the lowest three bands.
These indicate the degree to which each periodic function, at $k=0$, is required to represent a wavefunction with crystal momentum $k$ in band 1.
}
\label{fig:delta}
\end{figure}

The matrix elements of the transformation operator can be shown to be 
\begin{eqnarray}
U_{mn}(q,k) = \delta (k+A^{(g)}-q) \Delta_{mn} (k+A^{(g)},k) ,
\label{eqn:Umnqk}
\end{eqnarray}
where we have introduced the function
\begin{equation}
\Delta_{nm}(k_1,k_2) \equiv \langle u_{nk_1} | u_{mk_2} \rangle ,
\label{eqn:DeltaFunction}
\end{equation}
which is the overlap integral of the periodic parts of the Bloch wavefunctions within one unit cell. While the total wavefunctions, $| \phi_{nk} \rangle$, are all mutually orthogonal, this is not true for the periodic parts, $| u_{nk} \rangle$. Whenever $k_1=k_2$, the function $\Delta_{nm}(k_1,k_2)$ reduces to a Kronecker delta function, $\delta_{nm}$, but not otherwise. This is a statement of the fact that, at a particular value of the crystal momentum, $k=k_0$, the periodic functions form a complete orthonormal set, $u_{nk_0}(x)$, within the Hilbert space of a single unit cell. But a different choice of $k$ leads to a different set of functions, which will be mutually orthogonal, but need not be orthogonal to the functions of the first set. This can be seen in Fig. \ref{fig:delta}, which shows $| \Delta_{n,1}(0,k) |^2$, as a function of $k$ for different bands, $n$. At $k=0$, $\Delta_{11}(0,0) = 1$ while all other matrix elements are zero. However, as $k$ departs from zero, neighbouring bands become important (Fig. \ref{fig:delta}b), and as $k$ increases (over several Brillouin zones), higher-lying bands have the largest contributions (Fig. \ref{fig:delta}a). At a glance, this gives a sense of how many bands are required at the $\Gamma$-point to express the periodic part of the wavefunction at a different point, $k$, in reciprocal space. We note that these $\Delta$ functions have been discussed previously in the solid-state literature \cite{Marzari2012}. 

% We can now write the matrix elements of the projection operator for state $|\phi_{mq}\rangle$ as
% \begin{eqnarray}
% \left[ \Pi_{mq}^{(g)} \right]_{nn',k} 
% &\equiv& \langle \phi_{nk} | \Pi_{mq}^{(g)} | \phi_{n'k} \rangle \nonumber \\
% &=&\delta(k+A-q) \Delta_{mn}^*(q,k) \Delta_{mn'} (q,k).~~
% \label{eqn:ProjectionElements}
% \end{eqnarray}
From the Dirac delta function of Eq. \ref{eqn:Umnqk}, we immediately see that in the velocity gauge the operator that reports the population of the state with crystal momentum $q$ in band $m$ involves matrix elements with a different crystal momentum, $q-A^{(g)}$, and may involve matrix elements in all bands. This is why band populations \textit{appear} to be very different when calculated in the two gauges.

Finally, the matrix elements of the band projection operator, found by integrating the wavevector $q$ over one Brillouin zone, are
\begin{equation}
\left[ \hat{\Pi}_{m}^{(g)} \right]_{nn',kk'} 
= \Delta_{mn}^*(k+A^{(g)},k) \Delta_{mn'} (k+A^{(g)},k) \delta(k-k').~
\label{eqn:BandProjectionElements}
\end{equation}
The expectation value of this projection operator represents our proposed formulation of instantaneous band populations. Figure \ref{fig:3}a shows the conduction band population calculated using  Eq. \ref{eqn:BandProjectionElements} in the velocity gauge (dashed orange) and length gauge (filled blue). The two calculations overlap exactly, showing that this definition indeed gives a gauge-invariant value of the instantaneous band population.

The conduction band population exhibits sharp peaks at the vector potential zeros (near the peaks of the electric field), shown more clearly in Fig. \ref{fig:3}b. The sharpness of these peaks is in part due to the fact that the valence band is only occupied at a single value of $k$. A filled valence band will lead to broader transient peaks in the conduction band population. The maximum transient population transfer is about 5\%, much lower than the 40\% described by the previous velocity-gauge calculation of Fig. \ref{fig:2}a. We also note that, since the transformation operator for the length gauge is the identity operator, the length gauge calculations of Figs. \ref{fig:3}a and \ref{fig:2}a are identical. We will revisit this fact in Section \ref{sec:Discussion}.

\section{Interband and Intraband Currents}
\label{sec:currents}
Having considered the band populations during the pulse, and the closely related projection operators, we are in a position to revisit the question of interband and intraband currents. As before, we start by identifying a Hermitian operator, and then determine its gauge-dependent transformation. To do this, we again make use of projection operators to define the field-free operators
\begin{equation}
\hat{j}_{ra} = - \sum_{n} \hat{\Pi}_{n} \hat{p}_{\rm{kin}} \hat{\Pi}_{n}
\end{equation}
and
\begin{equation}
\hat{j}_{er} = - \sum_{n,n'\neq n} \hat{\Pi}_{n} \hat{p}_{\rm{kin}} \hat{\Pi}_{n'} .
\label{eqn:InterbandCurrent}
\end{equation}
The expectation values of these operators, in the absence of any applied fields, reproduce precisely the two summations shown in Eq. \ref{eqn:NaiveCurrent}. However, when a field is applied, a gauge must be chosen and the operators transformed appropriately. The transformation to the length gauge is, again, trivial and leaves the decomposition unchanged. However, when transforming the operator to the velocity gauge, we must transform not only the momentum operator, but also the projection operators. By simply dividing the current according to the terms shown in Eq. \ref{eqn:NaiveCurrent}, we perform the required transformations on the momentum operator ($\hat{p}^{(g)} = \hat{p}+A^{(g)}$) and the wavefunctions (Eq. \ref{eqn:WavefunctionTransformation}), but we fail to take into account the transformation of the projection operators.

The gauge-transformed intraband current operator is
\begin{equation}
\hat{j}_{ra}^{(g)} = - \sum_{m} \iint_{BZ} dq dq' \hat{\Pi}_{mq}^{(g)} (\hat{p}+A^{(g)}) \hat{\Pi}_{mq'}^{(g)} .
\label{eqn:IntrabandOperator}
\end{equation}
%where we may use Eq. \ref{eqn:ProjectionElements} to define the projection operators. 
Its matrix elements are 
\begin{eqnarray}
\left[ \hat{j}_{ra}^{(g)} \right]_{nn',kk'} 
&=&  - \sum_{mll'} \Delta^*_{mn} \Delta_{ml} 
\Delta^*_{ml'} \Delta_{mn'} \nonumber \\
&& \times \left[ p_{ll'}(k)  + \delta_{ll'} A^{(g)} \right] \delta(k-k') , 
\end{eqnarray}
where the arguments of the functions, $\Delta$, are understood to be $\Delta_{ab} \equiv \Delta_{ab}(k+A^{(g)},k)$.

\begin{figure}
\includegraphics{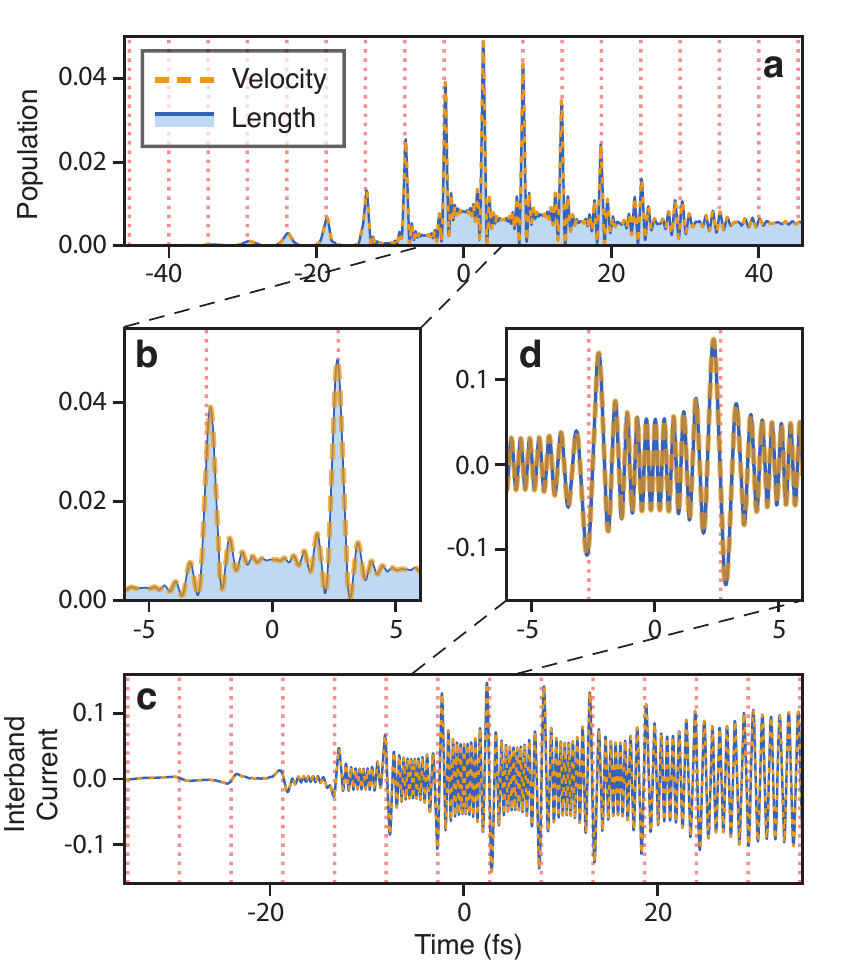}
\caption{(a) Conduction band population, defined using Eq.
\ref{eqn:BandProjectionElements}, as a function of time for velocity and length gauge,
which overlap. (b) Magnified view of the conduction band
population, showing transient peaks at the vector potential
zeros, as well as high-frequency oscillations. (c) Interband
current, defined using Eq. \ref{eqn:InterbandCurrentElements}, as a function of time for
velocity and length gauge. (d) Magnified view of interband
current shows that the two gauges agee within numerical
acuracy.}
\label{fig:3}
\end{figure}

Likewise, for the interband current, we have
\begin{equation}
\hat{j}_{er}^{(g)} = - \sum_{m, m'\neq m} \iint_{BZ} dq dq' \hat{\Pi}_{mq}^{(g)} (\hat{p}+A^{(g)}) \hat{\Pi}_{m'q'}^{(g)},
\label{eqn:InterbandOperator}
\end{equation}
whose matrix elements are
\begin{eqnarray}
\left[ \hat{j}_{er}^{(g)} \right]_{nn',kk'} 
&=& 
- \sum_{\substack{mm' \\ ll'}} 
\Delta^*_{mn} \Delta_{ml} \Delta^*_{m'l'} \Delta_{m'n'}  \nonumber \\
&& \times \left[ p_{ll'}(k) + \delta_{ll'} A^{(g)} \right] \delta(k-k') ,
%\Delta_{mn}^* \Delta_{ml}
%\left[ p_{ll'}(k) + \delta_{ll'} A^{(g)} \right] \Delta_{m'l'}^* \Delta_{m'n'}, \nonumber \\ ~
\label{eqn:InterbandCurrentElements}
\end{eqnarray}
where the summation omits terms with $m=m'$. Figures~\ref{fig:3}c and d show the interband current calculated using Equation \ref{eqn:InterbandCurrentElements} in the velocity gauge (dashed orange) and length gauge (solid blue). Here again, the agreement between the two calculations shows that our definition is gauge-invariant. 

\section{Discussion}
\label{sec:Discussion}
%\subsection*{Review and Fig 4}
We have proposed new formulations for the interband and intraband currents, as well as the instantaneous band populations. This was done by identifying Hermitian operators corresponding to each of these, and then applying a gauge-dependent unitary transformation. We have shown that the resulting operators give gauge-invariant predictions for these quantities, providing a possible resolution to the issue of gauge-dependence. In addition to this, our formulation allows all of these quantities to be computed from any gauge. In particular, the velocity gauge may present advantages regarding computation in some cases \cite{Han2010}. 

The harmonic spectrum for the strongly driven Mathieu potential, calculated entirely in the velocity gauge, is shown in Fig. 4. The total current (black line) displays the previously discussed double plateau. The difficulty in retrieving interband and intraband information was noted by Wu et al. \cite{Wu2015}, who suggested an approach involving projection onto Houston states. We agree with their approach and conclusions, however, since they project onto a time-dependent basis, they effectively change gauges and derive a new equation of motion for the wavefunction. We discuss this further in Appendix \ref{app:Houston}. Our approach allows us to consider the separation into interband (not shown) and intraband (red line) contributions, entirely in the velocity gauge. Furthermore, it is possible to make finer divisions of the current. For instance, we may define the interband current between bands one and two as $\hat{j}_{12} = \hat{\Pi}_{1} \hat{p}_{\rm{kin}} \hat{\Pi}_{2} + \rm{h.c.}$, whose spectrum is shown as the blue shaded region. This current accounts for the first plateau. Likewise, the summed interband currents from band one to bands three and four, $\hat{j}_{13}+\hat{j}_{14}$, shown as the pink shaded region, account for the second plateau. The relation of these plateaus to interband dynamics was previously argued based on the energy ranges at play and the intensity scaling \cite{Wu2016,Liu2017}, but here we can compute the interband dynamics directly even though our calculation is in the velocity gauge. 

\begin{figure}
\includegraphics{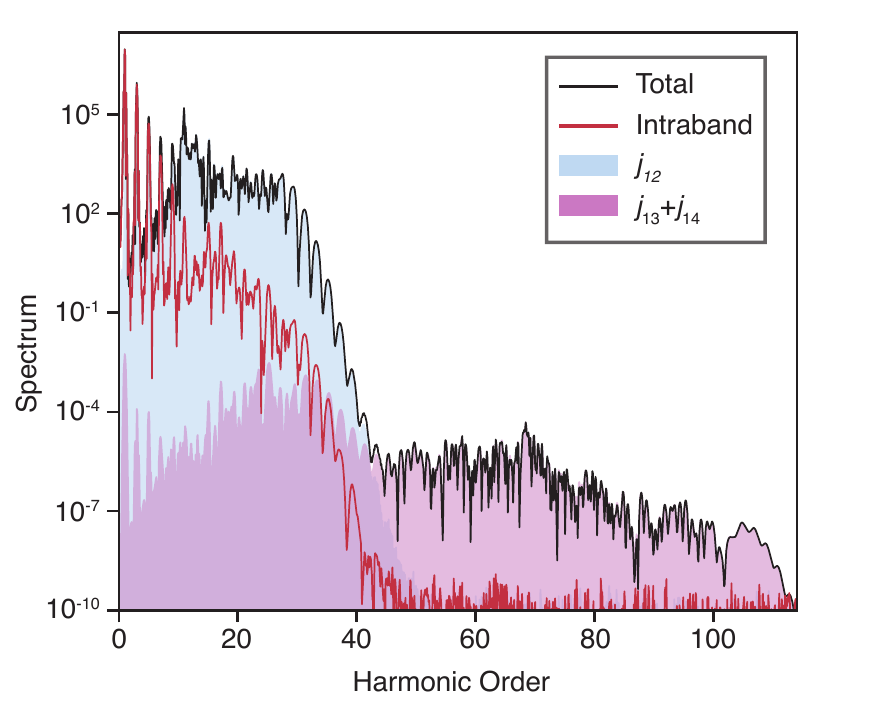}
\caption{Harmonic spectrum for the strongly driven
Mathieu potential. The black line shows the spectrum of
the total current. The red line shows that of the intraband
current which makes a negligible contribution for harmonics
above the seventh order. The blue shaded region
shows the spectrum of the interband current due to transitions
between bands one and two. Likewise, the pink
shaded region shows the spectrum of the combined interband
currents due to transitions between bands one and
thre, and bands one and four. All quantities shown are calculated in the velocity gauge.}
\label{fig:4}
\end{figure}

%\subsection*{Special place for length gauge}
It is worth noting that the gauge-invariant quantities that we defined coincide with the earlier problematic definitions in the length gauge (those of Section \ref{sec:NaiveDefinitions}). In our definitions, the field-free operators are transformed by the unitary operator 
%of Eq. \ref{eqn:UnitaryOperator}. 
$e^{iA^{(g)}x}$.
However, in the length gauge, since $A^{(l)}=0$, our approach leaves the field-free operator unchanged. But could a different choice have been made? In particular, could we not have chosen a definition such that the \textit{velocity} gauge operator corresponds to the field-free operator? Yes, such a choice could have been made, and the transformations of Eq. \ref{eqn:GeneralGaugeTransformation} would ensure gauge invariance. The resulting population operator in gauge $(g)$ would be
\begin{equation}
e^{-i[A^{(g)}-A^{(v)}] x} \hat{\Pi}_{mq} e^{i[A^{(g)}-A^{(v)}] x} ,
\label{eqn:AlternateProjection}
\end{equation}
to be compared with our proposed definition, Eq. \ref{eqn:InvariantBlochProjection}.
What, then, justifies the choice to associate the length gauge operator with the field-free operator? While we are not aware of any hard and fast justification for our choice, we suggest two possible arguments here, the first esthetic and the second an appeal to reasonableness. 

Esthetically, the alternate definition of Eq. \ref{eqn:AlternateProjection} defines the operator in gauge $(g)$ by making reference to the vector potential in that gauge and in gauge $(v)$, as well as referencing the field free operator. Thus, in any gauge, this alternate definition must always refer back to the velocity gauge. In contrast, our proposed definition, Eq. \ref{eqn:InvariantBlochProjection}, describes the operator in gauge $(g)$ only in terms of the field free operator and the vector potential in that gauge. Perhaps more importantly, our definitions give reasonable answers, matching the understanding described in the literature. Figure \ref{fig:4} shows that our gauge-invariant formulation explains the two plateaus in the harmonic spectrum as arising from different interband currents, consistent with previous arguments. A more straightforward example is the response of the system to a constant electric field. For a sufficiently weak field, one expects an adiabatic evolution of the wavefunction within a single band, together with Bloch oscillations in the intraband current. Appendix \ref{app:Bloch} shows that this behaviour is captured by our definitions, but not by alternative formulations like Eq. \ref{eqn:AlternateProjection}.

%\subsection*{Dressed states}
The questions we addressed in this article attempt to define instantaneous band-dependent quantities in the presence of a driving field. Yet the bands themselves are, in some sense, a field-free concept. A consideration of laser-dressed states is therefore an important part of this discussion. Laser-dressed states can be defined in various ways, either as instantaneous eigenstates of a Hamiltonian, which will depend on the chosen gauge, in a cycle-averaged (Floquet) way, or otherwise \cite{Higuchi2014,McDonald2017,Smirnova2006,Smirnova2007}. Here, we have not touched on these issues, aside from the Houston states, which are the instantaneous eigenstates of the velocity gauge Hamiltonian. In atomic and molecular systems, the dressed states have played an important role in resolving unphysical anomalies in calculations, which are often gauge-dependent \cite{Smirnova2006,Smirnova2007}. We may then expect such considerations to shed further light on questions of, for example, instantaneous state populations in a laser field. We leave a detailed consideration of dressed states, as they relate to gauge-dependence, for future study.

To conclude, we have proposed a formulation of instantaneous band populations, and interband and intraband currents which gives gauge invariant predictions. Our approach was to define Hermitian operators corresponding to these quantities, which can be done without making reference to a field or a gauge. These field-free operators are then transformed by the well-known unitary operator that connects gauges and ensures gauge invariance. We have demonstrated numerically that this gives identical results in the velocity and length gauges, and that these results are consistent with the community's use of these terms. The results give these important quantities a more rigorous definition, which, by removing the gauge-dependence, establishes that these are physically meaningful.

\section*{Acknowledgements}

For helpful discussions and comments, we would like to thank our colleagues, Michael Spanner, Thomas Brabec, Chris McDonald, Andr\'e Staudte, and Paul B. Corkum. This work was supported by the Vanier Canada Graduate Scholarship program (G.E.) and by the Air Force Office of Scientific Research Multidisciplinary University Research Initiative grant number FA9550-15-1-0037.
%We would like to thank ... for generous and stimulating discussions. We acknowledge the support of ... [Vanier scholarship] [Paul's grants?] [Anything else?].

%\subsection*{Conclusion}

\appendix

\section{Relation to Houston States}
\label{app:Houston}
% In previous studies of this same Mathieu potential, Wu \textit{et al.} employed a velocity gauge approach to describe the solid in analogy to a multi-level system [cite, cite]. In that work, the authors noted the difficulty in identifying interband and intraband currents from the velocity gauge, which is an expression of the same apparent gauge-dependence that we have addressed here. Their approach to resolving this issue was to consider a time-dependent basis set of Houston states. The Houston states are defined as 
% \begin{equation}
% a=a
% \end{equation}
Houston states have played an important role in the development of solid-state physics. They were originally proposed as approximate solutions of the length gauge Hamiltonian in a constant electric field \cite{Houston1940}. Later, they were employed by Krieger and Iafrate to find an analytical formulation of the TDSE in the velocity gauge \cite{Krieger1986}. This latter approach, starting from the velocity gauge, was used by Wu et al. to make the separation between interband and intraband currents \cite{Wu2015}. They showed that when the TDSE in the velocity gauge is solved in the basis of Houston states, the simple decomposition according to band indices gives sensible results. Indeed, it has long been noted that there is a close connection between the Houston state basis, and gauge transformations. In this Appendix, we address this connection.

In velocity gauge calculations, the Houston states are defined as
\begin{equation}
|\widetilde{\phi}_{nk_0} \rangle = e^{-iA^{(v)}x} | \phi_{nk(t)} \rangle,
\label{eqn:Houston}
\end{equation}
where $k(t) = k_0 + A^{(v)}$, and the vector potential is understood to be that of the velocity gauge. We make two important remarks on these states. First, we note that the crystal momentum (on the right side of the equation) acquires a time dependence which is a consequence of the acceleration theorem. Thus, the states $| \phi_{nk(t)} \rangle $ can be thought of as ``accelerated Bloch states''. They are eigenstates of $H_0$, albeit with a time-dependent crystal momentum. Second, the the phase factor which multiplies this state is precisely the unitary transformation that describes a gauge transformation, $U^{\dagger}$. Thus, the Houston states defined by Eq. \ref{eqn:Houston} can be thought of as the velocity gauge representation of accelerated Bloch states. Indeed, the original treatment by Houston did not include the transformation operator, $e^{-iA^{(v)}x}$, since it employed the length gauge.

Considering the TDSE in the Houston-state basis, numerically, one solves for the complex amplitudes of the basis states:
\begin{eqnarray}
\langle \widetilde{\phi}_{nk_0} | \psi^{(v)} \rangle 
&=& \left\{ \langle \phi_{nk(t)} | e^{iA^{(v)}x} \right\} \cdot \left\{ e^{-iA^{(v)}x} | \psi^{(l)} \rangle \right\} \nonumber \\
&=& \langle \phi_{nk(t)} | \psi^{(l)} \rangle .
\end{eqnarray}
That is to say the coefficients describing the Houston states in the velocity gauge are identical to those describing accelerated Bloch states in the length gauge. It is thus not surprising that the coupled differential equations describing these coefficients end up being essentially identical to the length gauge Schr\"{o}dinger equation. Indeed, one arrives at the same differential equations in the length gauge by using an accelerated frame \cite{Vampa2014}.

The approach of using a Houston-state basis does allow a decomposition into interband and intraband currents, however it involves a change in the differential equations employed, effectively requiring a return to the length gauge. In this sense, it does not directly address the issue of gauge-dependence. Also, it does not take advantage of the potential computational benefits of the velocity gauge. Our approach allows the TDSE to be solved entirely in the velocity gauge using the field-free basis, and still allows an unambiguous determination of interband and intraband currents. Furthermore, by defining quantities as Hermitian operators, we provide a more rigorous formulation which is equally valid in any gauge.

\section{Description of Bloch Oscillations in Velocity Gauge}
\label{app:Bloch}
The gauge-invariant definitions that we have proposed are chosen by first identifying the Hermitian operators corresponding to the observables of interest. This does not require us to think about a gauge, or even a field for that matter: currents or band populations are quantities that can be associated with the state of a system even in the absence of a field. Once these Hermitian operators are defined, gauge-invariance comes from the unitary transformation of Equation \ref{eqn:UnitaryOperator}. This is the thought process that leads to the definitions in Eqs. \ref{eqn:InvariantBlochProjection}, \ref{eqn:IntrabandOperator}, and \ref{eqn:InterbandOperator}. However, as noted above, the choice of a gauge-invariant definition based on a field-free operator is not unique. 
%In which gauge should the operator be the same as the field-free operator?
In this appendix, we use the simple case of Bloch oscillations to argue that our proposed definitions coincide with the commonly accepted and discussed behaviours of band populations and currents.

\begin{figure}[t!]
\includegraphics{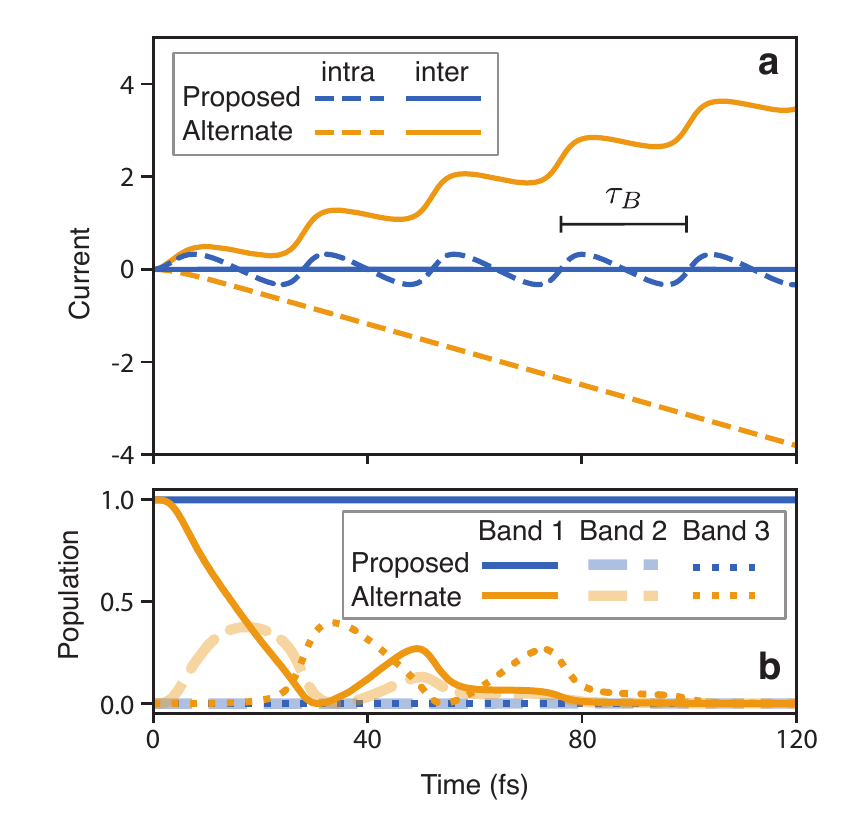}
\caption{
Dynamics of the Mathieu potential in a static electric field, comparing our proposed definitions (in blue) with the alternate definitions of Eqs. \ref{eqn:AlternateProjection}, \ref{eqn:AlternateIntraband}, and \ref{eqn:AlternateInterband} (in orange). (a) Interband and intraband currents. Bloch oscillations can be seen, and the Bloch period is shown. (b) Band populations as a function of time. In the proposed definition, the population remains in band 1 throughout, reflecting the adiabatic evolution of the state.
}
\label{fig:bloch}
\end{figure}

The definitions that we propose cause the length-gauge operator to be equal to the field free operator. An alternate definition for the instantaneous band population is described by Eq. \ref{eqn:AlternateProjection}, which instead causes the velocity gauge operator to coincide with the field-free operator. One can likewise put forward an alternate definition for the intraband current,
\begin{equation}
%\hat{j}_{ra} ,
e^{-i(A^{(g)}-A^{(v)})\hat{x}} \hat{j}_{ra}  e^{i(A^{(g)}-A^{(v)})\hat{x}} ,
\label{eqn:AlternateIntraband}
\end{equation}
and for the interband current,
\begin{equation}
e^{-i(A^{(g)}-A^{(v)})\hat{x}} \hat{j}_{er}  e^{i(A^{(g)}-A^{(v)})\hat{x}} .
\label{eqn:AlternateInterband}
\end{equation}
All of these alternate definitions also give gauge-invariant answers, like the ones we propose. However, the dynamics they describe do not match common expectations for an electron in a constant electric field. In a constant electric field, one expects an electron to evolve adiabatically within a single band (for sufficiently low field), undergoing Bloch oscillations.

Figure \ref{fig:bloch} shows the dynamics of the Mathieu potential in a constant field, comparing our proposed definitions to the alternate definitions of Eqs. \ref{eqn:AlternateProjection}, \ref{eqn:AlternateIntraband}, and \ref{eqn:AlternateInterband}. The constant field of $-7.93 \times 10^{-4}$ turns on exponentially, with a time constant of $4.1~\rm{fs}$, in order to avoid non-adiabatic excitation of higher bands. The initial condition is the same as in the previous calculations, a single electron in band 1 at $k=0$. 
%While the definitions are gauge invariant, we employ the velocity gauge, which requires us to use fifteen bands. 

In Fig. \ref{fig:bloch}a, the proposed definition of the intraband (dashed blue) clearly exhibits Bloch oscillations, while the interband (solid blue) current is zero, as expected. The current described by the alternate definitions (in orange) is noticeably different. Most importantly, the alternate definitions describe Bloch oscillations as an interband current, contrary to the community's understanding. The band populations described by our proposed definitions also match the common understanding. In Fig. \ref{fig:bloch}b, we see that the proposed definitions (blue) describe a population which remains entirely in band 1, with the other band populations remaining zero. This describes the adiabatic evolution of the electron within a single band. On the other hand, the alternate definition shows that the population moves from band 1 to bands 2 and 3, and eventually to higher-lying bands (not shown). The numerical results of Fig. \ref{fig:bloch} show that our proposed definitions give the correct results in a simple case where there is a consensus on the expected physical behaviour.

%\bibliography{main}

\begin{thebibliography}{36}
\bibitem [{NOT()}]{NOTEoperators}%
  \BibitemOpen
  \href@noop {} {}\bibinfo {note} {We note that the Hamiltonian operator does
  not transform like the operators of other Hermitian observable. Its
  transformation is found by substituting the transformed wavefunction into the
  TDSE.}\BibitemShut {Stop}%
\end{thebibliography}

\end{document}